\documentclass[prl,aps,reprint,nobibnotes,twocolumn]{revtex4-2}

\usepackage{graphicx}%
\usepackage[version=4]{mhchem}
\usepackage{seqsplit}

\begin{document}

\title{Embarking on a skyrmion Odyssey}

\author{Lei Shi}
\author{Zhiyuan Che}
\affiliation{Department of Physics, Fudan University, Shanghai, China}
	
\author{Yuri Kivshar}
\affiliation{Research School of Physics, Australian National University, Canberra, Australia}

\maketitle
\par \textit{Optical Skyrmions}, as an emergent cutting-edge topic in optics and photonics, extend the concept of non-singular topological defects to topological photonics, providing extra degrees of freedom for light-matter interaction manipulations, optical metrologies, optical communications, etc. \cite{R1}. The realization of artificial optical skyrmions was not due until 2018 \cite{R2,R3}, while the starting point of pursuits of optical skyrmions could date to Maxwellian and Kelvin's era, shown in Figure 1. The history of the rheological of skyrmions concept is somewhat similar to the homeward journey of Greek Myths hero Odyssey with twists and turns. The story goes all the way back to the old days when scientists discovered electromagnetism. Inspired by the fact of curl field nature of magnetism, Maxwell believed that electromagnetism should have rotational origin and proposed a model of ether vortices to derive the equations of electromagnetism \cite{R4}. After that, Lord Kelvin went a step further to propose an atomic model based on knots of swirling ethereal vortices immersed in Aether sea \cite{R5}. In the 1870s, there were huge debates over Kelvin’s vortex atoms model. Maxwell, a vortex atom enthusiast, promoted the model in his influential Encyclopedia Britannica article, 'Atom'. The opponent, like Boltzmann, said that the model lacks any proof of the validity of the equations. Along with the discovery of electrons and nuclei, the vortex atom hypothesis was finally abandoned, whereas the attractive features of those knots, including discreteness and immutability, have never been forgotten, and the idea of knots and knot invariants spawned a key modern physics conception, topological defects in the field theory. Around 60 years later, shown in Figure 1, the general interests of physicists changed from atoms to sub-atoms. The idea of knots returned to the stage, and it was employed by Skyrme to describe the nuclei \cite{R6,R7}. In Skyrme's picture, protons and neutrons are depicted as topological knots defects excitation in the three-component pion field, well known as \textit{skyrmions}. The number of knots twists, or knot invariants, equal to the number of nucleons in the nuclei. And by skyrmions, certain nuclei states had also correctly been predicted. Moreover, different from Kelvin's vortex atom hypothesis, the skyrmions in the nuclei are based on the nonlinear field theory with pion-pion interactions. And the nonlinear interaction physically guarantees that the skyrmions are stable under perturbations, in addition to the topological reason. Although it is accepted that the skyrmion is historically the first example of topological defects model, as the saying goes, the course of true love never did run smoothly. Along with the discovery of quarks, the skyrmion model was overlooked. Unexpected turns were associated with the rise of condensed matter physics. In condensed matter physics, a collective large number of atoms and electrons with fruitful symmetries, interactions as well as phases offer a platform to effectively construct various topological defects excitations, for example, vortices in superconductors, monopoles in spin ices and skyrmions in non-centrosymmetric magnetic systems \cite{R8,R9,R10}. Most importantly, condensed matter systems are sensitive to diversified external fields, leading to steerable manipulations of those topological defects with both versatility and precision. Therefore, since entering the condensed matter physics era, the concept of topology is not only used to explain natural matter, but also to control or even design matter, being the core in modern physics and related disciplines. Nevertheless, the skyrmion in condensed matter physics is still not a trouble-free journey. In the 1960s and 70s, it is believed that skyrmions are not expected to exist in most condensed matter systems due to the Hobart-Derrick theorem [9, 10]. Even in one of the most well-known papers of Nobel prize winners, Kosterlitz and Thouless, it is said that: "\textit{If we regard the direction of magnetization in space as giving a mapping of the space on to the surface of a unit sphere (actually it is exactly skyrmions), this invariant (skyrmions number) measures the number of times the map of the space encloses the sphere. This invariant is of no significance in statistical mechanics}" \cite{R11}. However, as marked in Figure 1, in 1989 A.N. Bogdanov and collaborators \cite{R12-1,R12-2} uncovered that magnetic materials with a broken inversion symmetry or, in other words, so-called noncentrosymmetric magnetic systems, could support \textit{magnetic skyrmions}. And it took another 20 years to realize the magnetic skyrmions experimentally \cite{R13-1,R13-2}. Since then, magnetic skyrmions became one of the hottest topics in condensed matter physics, constituting a promising new direction for data storage and spintronics \cite{R10,R14}. Of course, it is not the end of the Odyssey journey of Skyrmions. Ahead, there are still lots of challenges and opportunities, such as simultaneously increasing the stability and the transition temperature of magnetic skyrmions with nanometer size and realizations of skyrmions in other classical systems such as light. Recently, the storyline went to \textit{optical skyrmions}, as shown in Figure 1. 

\begin{figure*}[t!]
	\centering
	\includegraphics[width=0.96\textwidth]{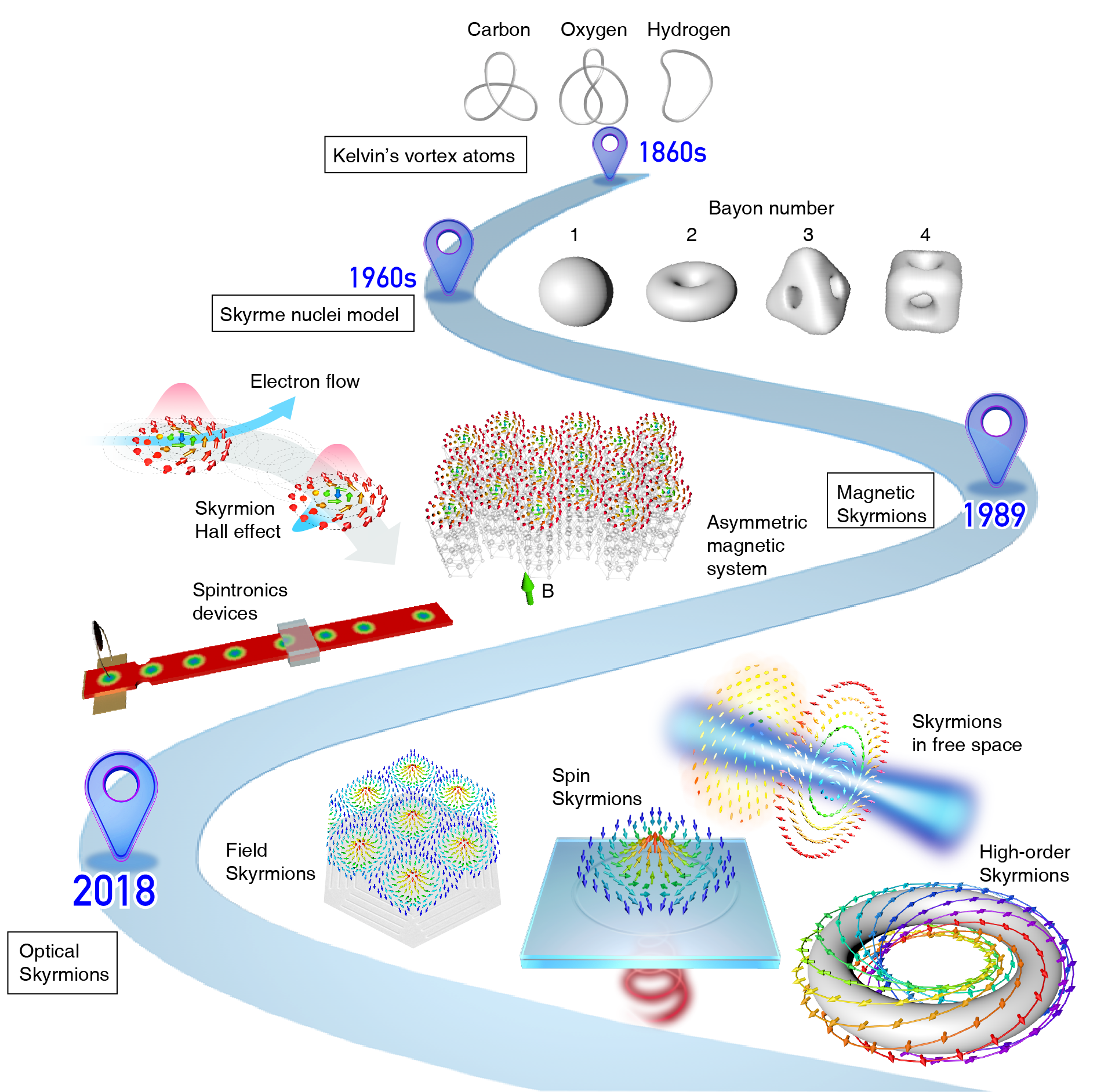}
	\caption{Navigating the skyrmion Odyssey: Tracing the key milestones from the inception of Maxwell’s and Kelvin’s theoretical concepts, through the progress in nuclear and condensed matter physics, to the recent emergence of optical skyrmions. }
	\label{fig:1}
\end{figure*}

In the recent review article published in Nature Photonics, Shen et al. \cite{R1} provided an extensive summary of the physics of optical skyrmions, covering nearly all the achievements since the first experimental realization of optical skyrmions. The review not only presents the fundamental theories and classifications of skyrmionic states, but also describes the generation and topological control of different kinds of skyrmions in evanescent, structured, and spatiotemporal optical fields through systematized literature reviewing. In addition, the authors discuss the generalized classes of optical nonsingular topological defects beyond skyrmions and outline emergent applications, promising future directions, and open challenges. As the concepts of optical skyrmion develops by leaps and bounds in the last 5 years, it is an excellent point in time to summarize this burgeoning research direction.

Shen et al. \cite{R1} start with the introduction of topological concept of optical skyrmions. Optical skyrmions are topological defects belonging to nontrivial homotopy classes in which various defined vector-like fields, such as spin, electrical field, polarizations states of light, in real space can be compactified into a hyper spherical surface of parameter space. Depending on the dimensionality of real space, optical skyrmions are divided into 2D baby skyrmions and 3D particle-like skyrmions. For 2D baby skyrmions, it can be intuitively understood as a vector field in 2D unwrapped from the vectors on a parameter sphere surface. Based on the orientations of the vector on the sphere surface, 2D baby skyrmions is classified by Néel, Bloch and anti-types. 2D Skyrmions can further generalized to merons (unwrapped from half sphere surface), bimerons and skyrmioniums (skyrmion with a radial multi-$\pi$-twist structure). For 3D particle-like skyrmions, it is frequently referred to hopfions, 3D vectorial fields with knots of closed iso-vector lines on a surface of tori. It is quite important to clarify the differences between skyrmions and vortices which is a frequently used word in topological photonics. Topologically, vortices correspond to homotopy class $\pi_1(S^1)$, while skyrmions correspond to $\pi_D (S^N)$, where $D\geq{N}$ and $D>1$. In other words, the appearance of vortices relies on singularities, or space with holes, while that of skyrmions do not. Therefore, skyrmions as nonsingular topological defects has enabled a novel topological approach for controlling electromagnetic fields.

Subsequently, Shen et al. \cite{R1} summarized various constructions of optical skyrmions. Light has multiple degree of freedoms, hence there is a multiplicity of combined degrees of freedoms to construct optical skyrmions. The vector field of optical skyrmions can be the electric field, spin angular momentum as well as polarization Stokes vector etc. And the space to unwrap hyper spherical surface in parameter space can be real space, space–time or momentum space. Real space electric field skyrmions have been realized in 2D planar optical systems supporting standing evanescent waves of surface plasmon or guided modes \cite{R2,R15}. And by further introducing winding phase wavefront, standing evanescent waves with orbital angular momentum support spin skyrmions or spin merons ensured by spin–orbit coupling in the evanescent field \cite{R2,R16}. 2D optical field and spin skyrmions both originated from longitudinal components of evanescent waves, except that field skyrmions are time-varying, while spin skyrmions is static. Propagating waves in free space can also be used to construct optical skyrmions, including polarization Stokes vector skyrmions and skyrmions pulse in space–time \cite{R17,R18,R19}. For Stokes skyrmions, it was well known as the full Poincaré structured vector beams with spatially variant polarization. Note that a skyrmionic beam indeed is a full Poincaré beam, while a full Poincaré beam may not be a skyrmionics one. The skyrmionic beam must satisfy the topological restrictions, namely a polarization pattern unwrapped from the Poincaré sphere. Importantly, particle-like 3D skyrmions, optical hopfions, also have been realized experimentally in free-space optics by superimposing of vectorial Laguerre-Gauss beams, realizing a mapping from 3D real space to a 3D generalized Poincaré hypersphere including phase \cite{R18}. Here, we would like to discuss one of the most intriguing issues of optical skyrmions, the constructions of optical skyrmions mainly based on linear superpositions of waves. However, for both skyrmions model in nuclei and magnetic systems, nonlinear interactions have been proved to be necessary. Why could optical skyrmions exist without nonlinear effect? From our point view, in most of the optical skyrmions works, there is indeed an artificial nonlinear operation, under which the vector field, supposed to be uneven in space, is normalized everywhere with unitary vector module.

Finally, Shen et al. \cite{R1} highlighted the important features and potential applications of optical skyrmions. Due to the topological nature of optical skyrmions, theoretical unlimited topological parameters of skyrmions have great advantages in information encoding and processing with multiplexed degrees of freedom. And fruitful vector textures of skyrmions offer a new platform to manipulate the interactions between light and topological quasiparticles in various system. Moreover, in contrast to normal light fields whose structures always focus on light intensity variations, limited to diffraction, the spatial and temporal size of vectorial skyrmion features in space time could be realized on a deep-subwavelength and on femtosecond scales respectively, as the vector variations such as spins are not subject to diffraction \cite{R3,R19}. It has potential applications in imaging, metrology and sensing beyond the diffraction limit. Apart from the above-mentioned important features of optical skyrmions, topological stability is always mentioned as one of features of optical skyrmions. Here, it should be noted that in the framework of Shen et al. review, the authors clarified the differences of topological stability and topological robustness of optical skyrmions. For topological stability, it corresponds to the stability of the field configuration during its evolution in time in the absence of external perturbations, which should be satisfied under the definition of skyrmions. For topological robustness, it corresponds to the protection of the field configuration during weak perturbations. Although it is still an open question deserving to be researched in detail, topological robustness of optical skyrmions has been proved to be strongly related to spin-orbital coupling \cite{R20}. As the space-polarization non-separable property of vector beams, scattering can easily distort both the local phase and polarizations, however, the global topology related to the non-separability of spatial mode and polarization is difficult to be modified \cite{R21,R22}.

Odyssey journey of skyrmions still continues, with many new papers are being published in different fields \cite{R23,R24}, For optical skyrmions, firstly, the conclusive investigations on full topological protection during propagation against weak and strong perturbations is urgent. Secondly, in most skyrmion systems, nonlinear effect is indispensable. Recently, optical pseudospin skyrmion Hall effect has been demonstrated in nonlinear media \cite{R25}. Introducing nonlinear optics in optical skyrmions to reveal novel topological selection rule of mode interaction is another important topic. Thirdly, the pursuit of the existence of high order skyrmions in high dimensions, such as 3D hopfion and hopfion links, which are difficult to reach otherwise, is of vital importance \cite{R18,R26}. Meanwhile, diverse kinds of optical vectors, such as field vectors, spin vectors, energy flow vectors etc. and physical domains, such as space, momentum space \cite{R27}, space-time and even synthetic dimensions, can be used to do the topological mapping. In principle, there is unlimited scope, unlimited topologies, unlimited optical systems for us to explore optical topological states, enriching toolkits in optics and photonics. Finally, smoking gun applications of skyrmions are still cloaked in mist. The list of outlooks shown here is by far not complete. The Odyssey journey ahead may be long and arduous, but with sustained actions, we will eventually understand great opportunities brought by optical skyrmions, and bear witness to fuel rapid advances in the future.







%

\end{document}